\documentstyle[sprocl,amssymb,psfig]{article}

\def\be{\begin{equation}}
\def\ee{\end{equation}}
\def\bea{\begin{eqnarray}}
\def\eea{\end{eqnarray}}

\begin{document}

\title{Thermodynamic Properties of Correlated \\
Strongly Degenerate Plasmas}
\author{V.S.~Filinov, P.R.~Levashov, V.E.~Fortov}
\address{Russian Academy of Sciences, High Energy Density Research Center\\
Izhorskaya street 13-19, Moscow 127412, Russia}

\author{M.~Bonitz}

\address{Universit\"at Rostock, Fachbereich Physik\\
Universit\"atsplatz 3, 18051 Rostock, FRG \\
E-mail: micha@elde.mpg.uni-rostock.de}

\maketitle\abstracts{An efficient numerical approach to
equilibrium properties of strongly coupled systems which include a subsystem
of fermionic quantum particles and a subsystem of classical particles is
presented. It uses an improved path integral representation of the
many-particle density operator and allows to describe situations of
{\em strong coupling and strong degeneracy}, where analytical theories fail. A
novel numerical method is developed, which allows to treat degenerate
systems with full account of the spin scatistics. Numerical results for
thermodynamic properties such as internal energy, pressure and pair
correlation functions are presented over a wide range of degeneracy
parameter.}

\section{Introduction}\label{sec:intro}
Thermodynamic properties of correlated Fermi systems continue to
attract the interest of researches in many fields, including
plasmas, condensed matter and astrophysics.
All present theories have practical difficulties. Quantum kinetic
methods, including the Kadanoff-Baym equations
\cite{kadanoff-baym,green-book}, easily handle quantum
and spin effects, but they become extremely
complicated when correlation effects can not be treated perturbatively.
It is, therefore, important to consider alternative approaches, such as
Monte Carlo simulations \cite{zamalin,binder96}, as they allow for an
efficient treatment
of strong coupling phenomena. However, they still have difficulties in
incorporating quantum and spin statistics effect.
In particular, there has been remarkable progress in applying path
integral quantum Monte Carlo (PIMC) techniques to Bose systems
\cite{ceperley95rmp} and Coulomb systems, see e.g.
\cite{binder96,pierleoni-etal.94} and Ref.~\cite{ceperley95} for
an overview. Nevertheless, there has still been one major obstacle preventing
efficient modelling of Fermi systems -- the so-called sign problem
resulting from summation over all permuations of the
density matrix. In this paper, whe present a different path integral
representation in which no sign problem appears.

\section{Path-Integral Monte Carlo Methods}
As is well known the thermodynamic properties of a quantum system of $N$
particles are fully
determined by the partition function Z and, consequently, by the density
matrix
\begin{eqnarray}
Z = \int\limits_{V} dq \,\rho(q,0;q,\beta), \qquad
\rho(q,0;q',\beta) = \langle q|{\hat \rho}|q'\rangle ,
\label{z-rho}
\end{eqnarray}
where ${\hat \rho}=\exp\{-\beta {\hat H}\}$,
$\beta=1/kT$, and
$q$ comprises the coordinates of all particles,
$q \equiv \{{\bf q}_1, {\bf q}_2,..., {\bf q}_N\}$. With an analytical
expression for the density matrix given, one can use e.g. Monte Carlo methods
\cite{zamalin,binder96} to evaluate the partition
function and thermodynamic quantities. However, for a quantum system, $\rho$ is, in
general, not known, but can be constructed from its known high-temperature
limit by means of a decomposition into $n+1$ factors, each of which
corresponds to the density matrix at an $n+1$ times higher temperature
\cite{feynman-hibbs,zamalin,binder96,ceperley95,berne98},
\begin{eqnarray}
\rho(q,0;q',\beta) = \int\limits_{V} dq^{(1)} \dots dq^{(n)} \,
\rho(q,0;q^{(1)},\tau^{(1)}) \times \dots
\times \rho(q^{(n)},\tau^{(n)};q',\beta), \label{rho-fac}
\end{eqnarray}
where $\tau^{(i+1)}-\tau^{(i)}=\Delta \beta = \beta/(n+1)$.
Eq.~(\ref{rho-fac}) lets one view the N-particle state as a loop consisting of
n+1 vertices (``beads'') located at intermediate coordinates $q^{(i)}$
which is closed, i.e. $q^{(n+1)}\equiv q^{(1)}$, due to the trace in
Eq.~(\ref{z-rho}).
For quantum systems of bosons (fermions), furthermore, the spin statistics
has to be taken into account requiring to perform an (anti-)symmetrization of
the density matrix, after which Eq. (\ref{rho-fac}) obtains the form
(for simplicity, the spin variables are not written explicitly)
\begin{eqnarray} \rho(q,0;q',\beta) &=& \frac{1}{N!} \sum_{P} (\pm
1)^{\kappa_P} \int\limits_{V} dq^{(1)} \dots dq^{(n)} \,
\rho(q,0;q^{(1)},\tau^{(1)}) \times \dots
\nonumber \\
&& \qquad \qquad \qquad \qquad \times \rho(q^{(n)},\tau^{(n)};P q',\beta),
\label{rho-sym}
\end{eqnarray}
where $P$ denotes an arbitrary $N$-particle permutation of the particle
coordinates, and $\kappa_P$ is its parity.

The unknown density matrix (\ref{rho-sym}) is efficiently computed by
approximating the $\rho$'s on the r.h.s. of Eq.~(\ref{rho-sym}) by the
high-temperature density matrix $\rho^{hT}$. For potentials which are
bounded from below the simplest choice is
\cite{zamalin,binder96,ceperley95,berne98}
\begin{eqnarray}
\rho^{hT}(q,\tau;q',\tau') &=& \rho_0(q,\tau;q',\tau') \,
e^{-(\tau'-\tau)U(q)},
\nonumber\\
\rho_0(q,\tau;q',\tau')&=& \frac{1}{\lambda_{\Delta}^{3N}}\,
e^{- \pi |q-q'|^2/\lambda_{\Delta}^{2} },
\label{rho-ht}
\end{eqnarray}
where $\rho_0$ is the free-particle density matrix, $U$ is the potential
energy, and $\lambda_{\Delta}$ is the thermal DeBroglie wave length
corresponding to the higher temperature
 $kT'=1/\Delta \beta$,
$\lambda_{\Delta}^{2} \equiv 2\pi\hbar^2(\tau-\tau')/m$. It is well known
that for $n\rightarrow \infty$, $\rho^{hT}$ converges to the exact
density matrix  $\rho$. Notice that the factorization of $\rho^{hT}$ into
a kinetic and potential energy term, Eq.~(\ref{rho-ht}), is an approximation
also. The error made thereby is of the order of the variation of $U$ on the
spatial scale $\lambda_{\Delta}$ and thus vanishes with $n\rightarrow \infty$.
This holds also for an repulsive Coulomb potential, whereas the
attractive electron-ion interaction has to be represented by a bounded from
below effective pair potential \cite{green-book,zamalin,filinov76}.

\section{Thermodynamic quantities of dense quantum plasmas}\label{sec:td}
The sign problem is solved by a simple transform of the intermediate
electron coordinates \cite{filinov-etal.99prl}. To this end, we rewrite
the partition function
(\ref{z-rho}), now explicitly including the (classical) ion component,
$N_e=N_i=N$,
\begin{eqnarray}
Z(N,V,\beta) &=&
\frac{Q(N_e,N_i,\beta)}{N_e!N_i! \,\lambda_i^{3N_i}\lambda_{\Delta}^{3N_e}},
\nonumber\\
\mbox{with} \qquad
Q(N_e,N_i,\beta) &=& \int\limits_V dq \,dr \,d\xi
\,\rho(q,[r],\beta). \label{q-def}
\end{eqnarray}
Here, the notation $q$ is retained for the ions. $[r]$ summarizes the
electron coordinates with $r$ denoting the coordinates at the beginning
of the loop and $\xi^{(1)}, \dots \xi^{(n)}$ the dimensionless distances
between neighboring vertices. Thus, explicitly,
$[r]=[r, r+\lambda_{\Delta}\xi^{(1)},
r+\lambda_{\Delta}(\xi^{(1)}+\xi^{(2)}), \dots]$, and $q, r, \xi^{(i)}$
each are $3N$-dimensional vectors. For the density matrix in
Eq.~(\ref{q-def}) we have \cite{zamalin,filinov76}
\begin{eqnarray}
&&\rho(q,[r],\beta) = \sum\limits_{s=0}^N \rho_s(q,[r],\beta)
= \sum\limits_{s=0}^N \frac{C^s_N}{2^N}\,
e^{-\beta U(q,[r],\beta)} \prod\limits_{l=1}^n
\prod\limits_{p=1}^N \phi^l_{pp}
{\rm det} \,|\phi^{n,1}_{ab}|_s,
\nonumber\\
&&\mbox{where}\quad
 U(q,[r],\beta) = U^i(q) +
\frac{1}{n+1 }\sum\limits_{l=0}^n \left\{
U^e_l([r],\beta) + U^{ei}_l(q,[r],\beta)
\right\},
\label{rho-def}
\end{eqnarray}
and $U^i$ denotes the ion-ion interaction energy, whereas $U^e_l$
denotes the interaction energy corresponding to the electronic
vertex ``l'', and $U^{ei}_l$ means the analogous electron-ion
contribution. Furthermore, $\phi^l_{pp}\equiv \exp[-\pi |\xi^l_p|^2]$
arises from the kinetic energy density matrix $\rho_0$, while
$\phi_{ab}$ is the exchange matrix the elements of which are given by
$\phi^{n,1}_{ab}\equiv \exp[-\pi |(r_a-r_b)/\lambda_{\Delta}+
\sum_{k=1}^{n}\xi^k_a|^2]$. The determinant arises from the sum over
the permutations, cf. Eq. (\ref{rho-sym}).
Finally, $s$ is the number of electrons having
the same spin projection. Due to the spin variables, the exchange matrix
contains two zero submatrices which are related to electrons with
opposite spin projection.

As an example of applying our method to thermodynamic properties,
we provide the result for the equation of state,
$\beta p = \partial {\rm ln} Q / \partial V =
[\alpha \partial {\rm ln} Q
/ 3 V \partial \alpha]_{\alpha=1}$,
\begin{eqnarray}
\frac{\beta p V}{2N} &=& 1 + \frac{1}{6NQ}\sum\limits_{s=0}^N
\int dq \, dr \, d\xi \,\rho_s(q,[r],\beta) \,\Bigg\{
\sum_{p<t}^{N_i} \frac{\beta e^2}{|q_{pt}|} +
\sum_{p<t}^{N_e} \frac{\beta e^2}{(n+1)|r_{pt}|}
\nonumber\\
&-&  \sum_{p=1}^{N_i}\sum_{t=1}^{N_e} \frac{|x_{pt}|}{n+1}
\frac{\partial \beta \Phi^{ie}}{\partial |x_{pt}|}
+ \sum_{p=1}^{n}\sum_{p<t}^{N_e}
\frac{\beta e^2 \langle r^l_{pt}|r_{pt}\rangle }{(n+1)|r^l_{pt}|^3}
\nonumber\\
&-& \sum_{l=1}^{n}\sum_{p=1}^{N_i}\sum_{t=1}^{N_e}
\frac{\langle x^l_{pt}|x_{pt}\rangle}{(n+1)|x^l_{pt}|}
\frac{\partial \beta \Phi^{ie}}{\partial |x^l_{pt}|}
\,+\,\frac{\alpha}{{\rm det} |\phi^{n,1}_{pt}|_s}
\frac{\partial{\rm det} | \phi^{n,1}_{pt} |_s}{\partial \alpha}
\Bigg\}.
\label{eos}
\end{eqnarray}
Here, $\Phi^{ie}$ is the effective electron-ion pair potential,
$\alpha$ is a scaling parameter,
$\langle \dots | \dots \rangle$ denotes the scalar product, and
$q_{pt}, \xi_{pt}$, $r_{pt}$ and $x_{pt}$ are differences of two
coordinate vectors:
$q_{pt}\equiv q_p-q_t$, $r_{pt}\equiv r_p-r_t$,
$\xi^k_{pt} \equiv \xi^k_{p}-\xi^k_{t}$,
$r^l_{pt}\equiv r^l_{p}-r^l_{t}+\lambda_{\Delta}\sum_{k=1}^{l}\xi^k_{pt}$,
and $x^l_{pt}\equiv r^l_{p}-q_{t}+\lambda_{\Delta}\sum_{k=1}^{l}\xi^k_{p}$.
We underline that introducing the dimensionless coordinate components
$\xi$ removes a substantial part of the complicated temperature and volume
dependence from the density
\begin{eqnarray}
\centerline{
\psfig{file=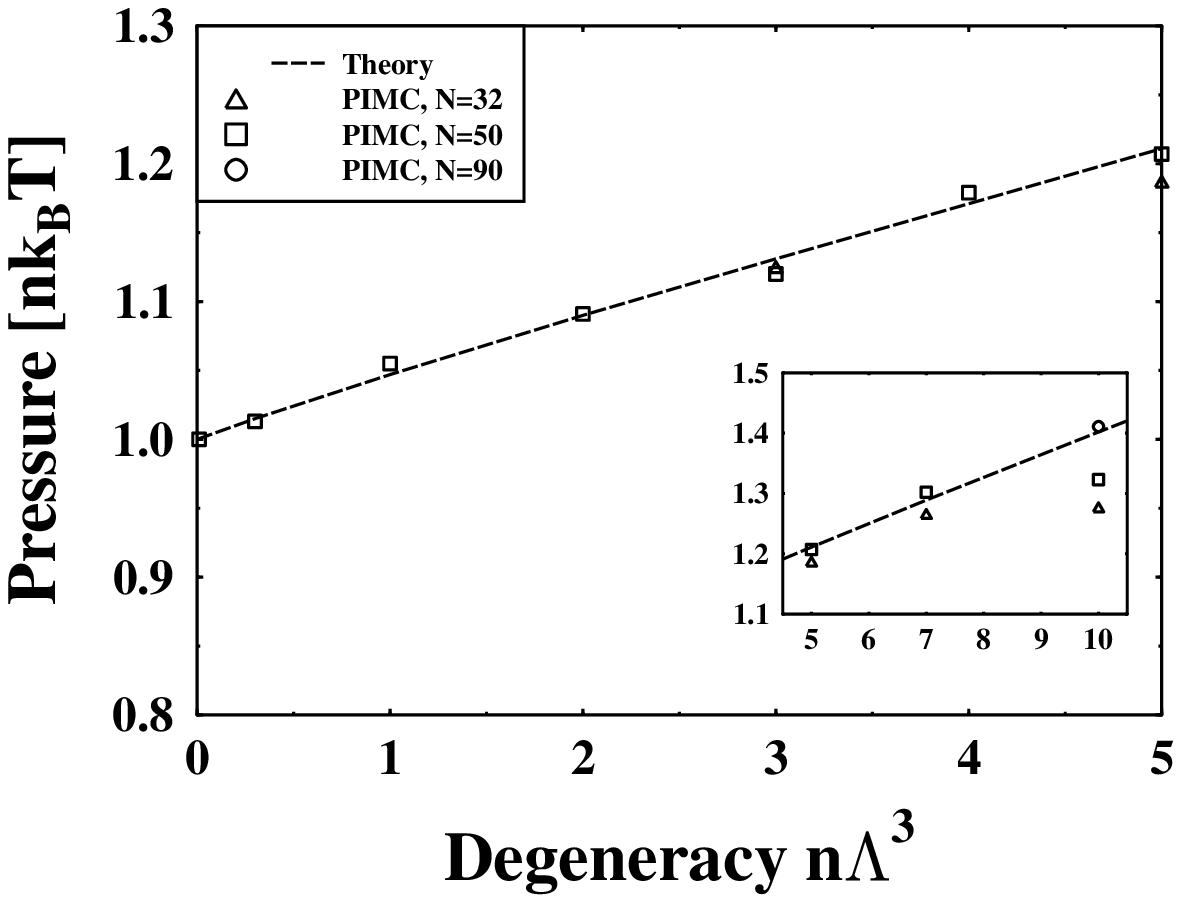,height=6cm,width=10cm,angle=-0}}
\nonumber
\end{eqnarray}

\vspace{-0.6cm}
\noindent \footnotesize
Figure~1:
Pressure of a mixture of ideal electrons and
protons. Spin statistics of the electrons is fully included, protons are
treated classically.

\vspace{0.3cm}
\noindent\normalsize
matrix (\ref{rho-def}). This turns out to be
crucial in the computation of thermodynamic quantities as they involve
derivatives of the partition function with respect to $V$ or $\beta$.
Furthermore, the anti-symmetrization propblem is reduced to computing
the determinant of $\phi_{pt}$, and no explicit summation over
permuations with alternating sign is required.
In fact, for each fixed particle configuration, the absolute value of each
sum in curly brackets in Eq.~(\ref{eos}) is bounded
as $n\rightarrow \infty$, which is sufficient to eliminate the sign
problem completely. This enables us to evaluate thermodynamic properties
of quantum plasmas without further approximations on the density matrix
and the integration region. For the numerical calculations, we use the standard
Metropolis Monte Carlo procedure in which the probability of sampling
configurations is proportional to $|\rho_s|$, while the sign of $\rho_s$ is
included into the weight function of the particle configuration.
Expressions similar to (\ref{eos}) are readily derived for other
thermodynamic quantities  and, along with more details, will be given
elsewhere.

\section{Numerical results}\label{sec:res}
Using the above results, we have performed a series of calculations for
a two-component electron-proton plasma. Notice that this
first-principle treatment of the elementary plasma particles (physical
picture) does fully include bound states. Of course, the quality of
describing their properties depends on the potential $\Phi^{ie}$ for
which we chose an effective quantum pair potential originally derived by
Kelbg which is the  exact high-temperature limit, e.g. \cite{green-book}.
Thus, by increasing the number of electronic  vertices $n$, in
principle, any desired accuracy can be achieved.
To test the quality of the reproduction of quantum and spin statistics
effects, we first  consider a mixture of ideal electrons and protons for which the
thermodynamic quantities are known analytically, e.g. \cite{green-book}.
\vspace{-0.5cm}
\begin{eqnarray}
\centerline{
\psfig{file=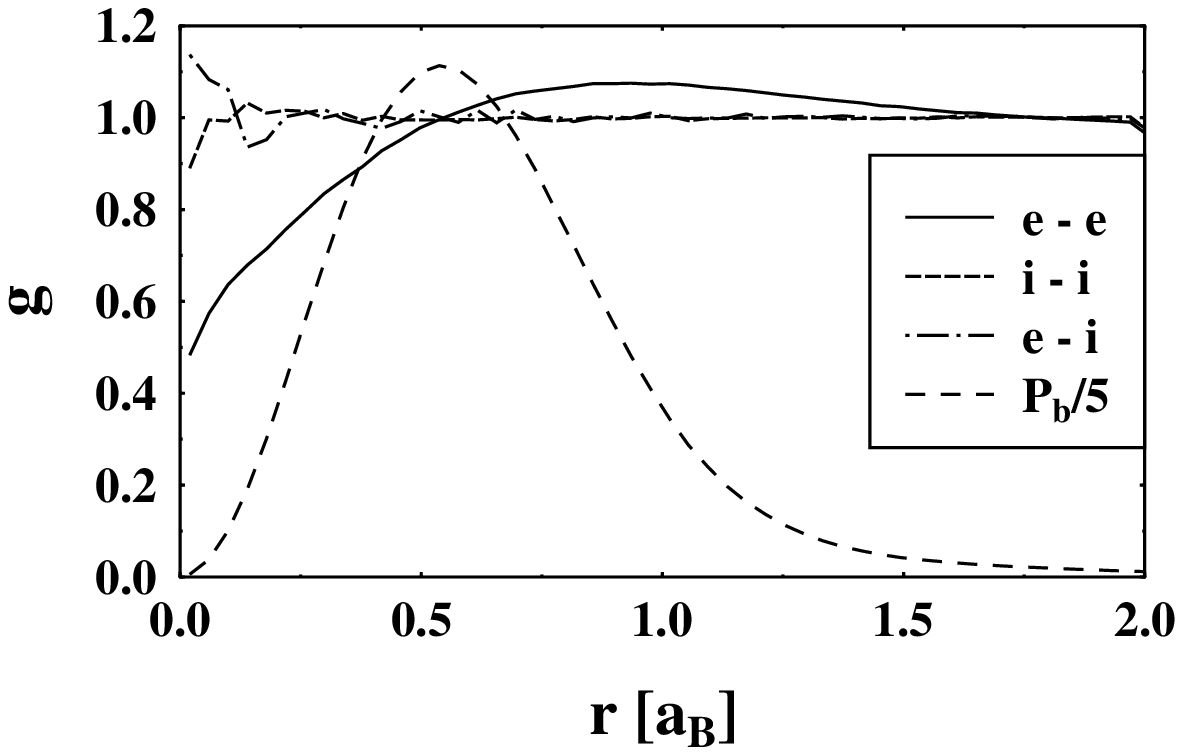,height=6cm,width=11cm}}
\nonumber
\end{eqnarray}

\vspace{-0.4cm}
\noindent \footnotesize
Figure~2: Pair correlation functions for
non-interacting electrons and protons at $n\Lambda^3=4$.
The decrease of
$g_{ee}$ for small distances reflects the Pauli principle.
Also shown is the vertex separation distribution $P_b$.

\vspace{0.3cm}
\normalsize \noindent
Fig.~1 shows our numerical results for the pressure together with the
theoretical curve. The agreement, up to values of the degeneracy parameter
$\chi\equiv n\lambda^3$ as large as 5 is remarkable. Even with only
$N=16$ electrons and protons deviations are rather small.
One clearly sees that increasing the number of particles improves the
numerical  results systematically.
\vspace{-0.4cm}
\begin{eqnarray}
\centerline{
\psfig{file=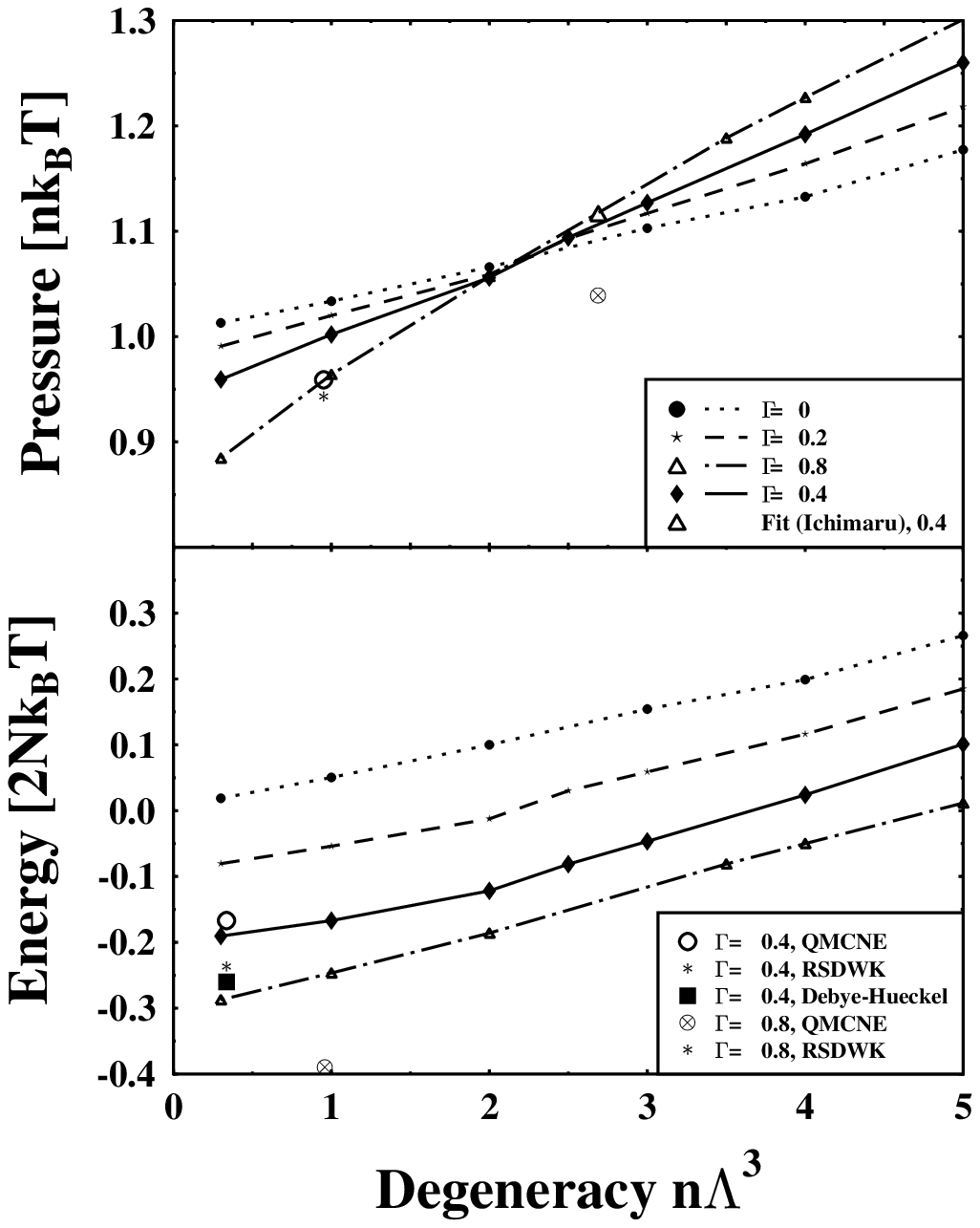,height=12cm,width=11cm}}
\nonumber
\end{eqnarray}

\vspace{-1.4cm}
\noindent \footnotesize
Figure~3: Pressure (upper figure) and total energy $-3NkT$
(lower figure) of an electron-proton plasma for different values of the classical
coupling parameter $\Gamma$. For comparison, results of other groups are
shown: QMCNE - quantum MC without exchange, RSDWK - analytical model of
Rieman et al. \cite{schla-data}.

\vspace{0.3cm}
\normalsize
Fig.~2 shows the correspondig
e-e, i-i and e-i  pair correlation functions. As expected, the functions
$g_{ei}$ and $g_{ii}$ are identical to one. The fluctuations at
small distances reflect the maximal statistical error of the MC simulation
and rapidly decay with increasing distance. The decay at large distances is
a consequence of the periodic
boundary conditions in the MC simulation.
In contrast, the electron-electron correlation function decays
at small distances reaching 0.5 at $r=0$ which is the expected result for
particles with spin $1/2$. Notice further the maximum of $g_{ee}$ which
appears around the thermal wavelength $\lambda_e$ and reflects weak ordering
due to Fermi repulsion. To characterize the strength of quantum effects, we
also show the probability distribution of distances between neighboring
vertices on the electron loops $P_b$. It has a maximum on a distance
of the order of $\lambda_{\Delta}/\sqrt{\pi}$.

\vspace{-0.5cm}
\begin{eqnarray}
\centerline{
\psfig{file=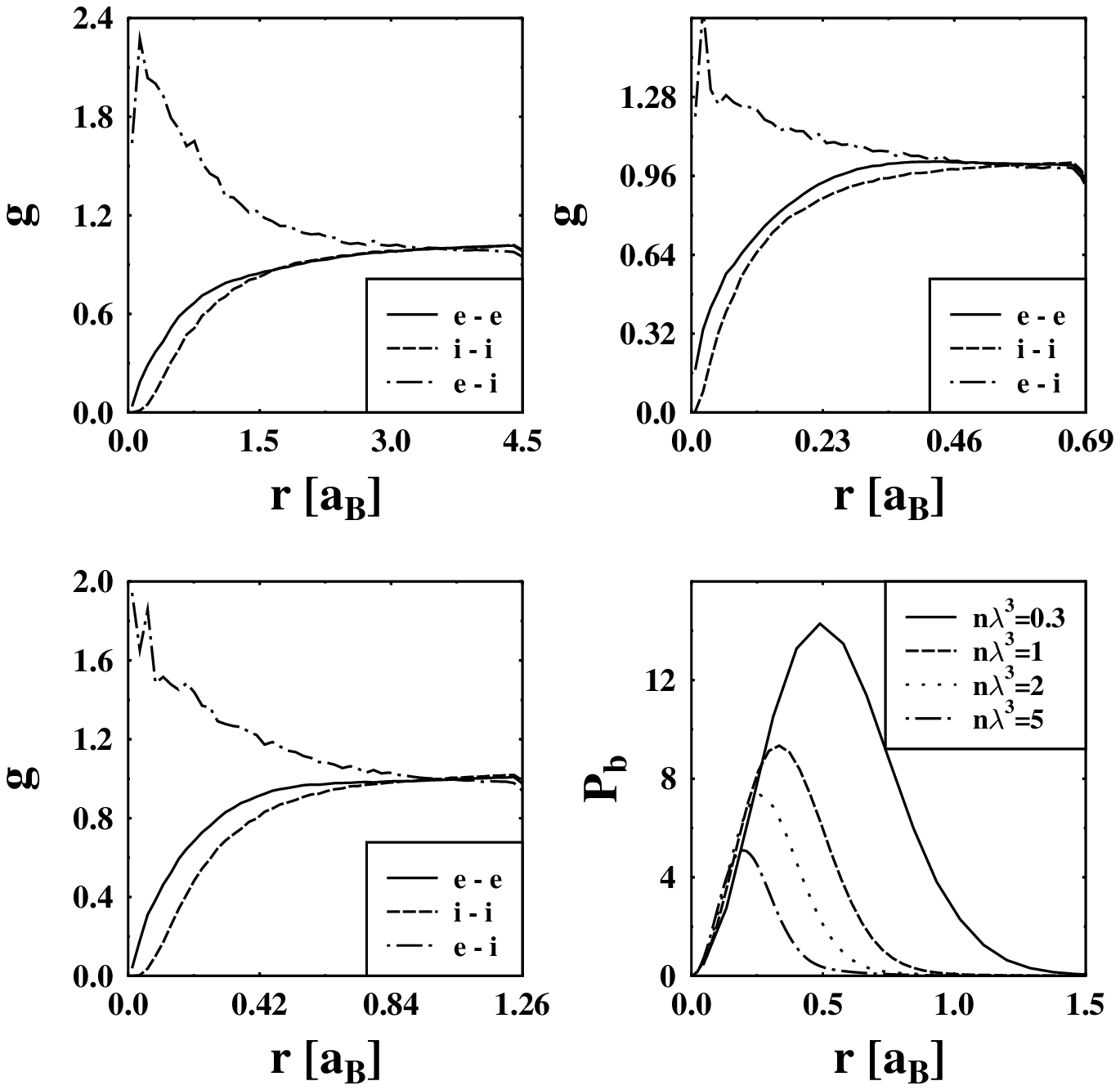,height=9cm,width=10cm}}
\nonumber
\end{eqnarray}

\vspace{0.cm}
\noindent \footnotesize
Figure~4: Pair correlation functions for a
correlated electron-proton plasma with $\Gamma=0.4$ and different
values of the degeneracy parameter. Notice the different length scales.
Bottom right figure shows the vertex separation distribution $P_b$
(arbitrary units, see text).

\vspace{0.3cm}
\normalsize
Let us now turn to the case of interacting electrons and protons. We have
performed a series of calculations in which the classical coupling parameter
$\Gamma=(4\pi n_e/3)^{1/3}e^2/4\pi \epsilon_0 kT$ was kept constant
while the degeneracy was varied. The results for the pressure and
energy are presented in Fig.~3. One can see that for weak coupling and
small degeneracy parameters, $\chi<0.5$, exchange effects are small, and QMC
simulations  without exchange (open circles) are close to our results.
However, with increasing $\chi$ and $\Gamma$, the deviations are growing
rapidly. Even stronger are the discrepancies with analytical theories
which are constructed as perturbations expansions, and thus are limited
to small values of $\chi$ and $\Gamma$. Most strikingly is the decrease
of the energy as a function of $\chi$ predicted by the analytical
models and QMC without exchange, which is in
contrast to our results which show an increase for all values of
$\Gamma$.
Finally, Fig.~4 shows the pair correlation functions for
intermediate coupling, $\Gamma=0.4$ for various degrees of degeneracy.
Now, due to Coulomb repulsion, at small distances $g_{ee}$ and
$g_{ii}$ decay to zero. However, the decay of $g_{ee}$ is essentially
different from  that of the proton-proton function, despite the identical
Coulomb force. The reason  are quantum exchange and tunneling effects in
the electron subsystem  which compete with the Coulomb repulsion.


\section*{Acknowledgments}
We acknowledge support from the Deutsche Forschungsgemeinschaft
(Mercator-Programm) for VSF and stimulating discussions with
B. Bernu, D. Ceperley, D. Kremp and M. Schlanges.



\section*{References}
\begin{thebibliography}{99}

\bibitem{kadanoff-baym}
L.P. Kadanoff and G. Baym,
{\em Quantum Statistical Mechanics},
Addison-Wesley Publ. Co. Inc., 2nd ed., 1989.

\bibitem{green-book} W.D. Kraeft, D. Kremp, W. Ebeling, and G. R\"opke,
{\em Quantum Statistics of Charged Particle Systems}, Akademie-Verlag
Berlin 1986

\bibitem{zamalin} V.M. Zamalin,  G.E.Norman, and V.S. Filinov,
{\em The Monte Carlo Method in Statistical Thermodynamics},
Nauka, Moscow 1977 (in Russian).

\bibitem{binder96} {\em The Monte Carlo and Molecular Dynamics of
Condensed Matter Systems}, K. Binder and G. Cicotti (eds.),
SIF, Bologna 1996

\bibitem{ceperley95rmp} D. M. Ceperley,
Rev. Mod. Phys. {\bf 65}, 279 (1995)

\bibitem{feynman-hibbs} R.P. Feynman, and A.R. Hibbs,
{\em Quantum mechanics and path integrals}, McGraw-Hill, New York 1965

\bibitem{pierleoni-etal.94} C. Pierleoni, B. Bernu, D.M. Ceperley, and W.R. Magro,
Phys. Rev. Lett. {\bf 73}, 2145 (1994)

\bibitem{ceperley95} D.M. Ceperley, in Ref. \cite{binder96}, pp. 447-482

\bibitem{berne98} {\em Classical and Quantum Dynamics of Condensed Phase
Simulation}, B.J.Berne, G.Ciccotti and D.F.Coker eds., World Scientific,
Singapore 1998

\bibitem{filinov76} V.S. Filinov,  High Temp. {\bf 13}, 1065 (1975) and
{\bf 14}, 225 (1976)

\bibitem{filinov-etal.99prl} V.S. Filinov, and M. Bonitz, subm. to Phys. Rev.
Lett.

\bibitem{schla-data} J. Riemann, M. Schlanges, H.E. DeWitt, and W.D. Kraeft,
Proceedings of the International Conference on Strongly Coupled Plasmas,
W.D. Kraeft and M. Schlanges (eds.), World Scientific, Singapore 1996, p. 82
\end{thebibliography}
\end{document}